\documentclass[epjCONF]{svjour}
\usepackage{graphicx}
\usepackage[varg]{txfonts} 
\usepackage[latin1]{inputenc}
\usepackage[english]{babel}
\usepackage{amssymb,amsfonts,amstext}
\usepackage{subeqn}
\usepackage{microtype}
\session-title{%
19$^{\textnormal{\footnotesize th}}$ International %
IUPAP Conference on Few-Body Problems in Physics%
}
\begin{document}
\title{%
Regge-plus-resonance predictions for charged-kaon photoproduction from the deuteron
}%
\author{%
P.~Vancraeyveld\thanks{\email{Pieter.Vancraeyveld@UGent.be}} 
\and %
L.~De~Cruz 
\and %
J.~Ryckebusch
\and %
T.~Van~Cauteren
}
\institute{%
Department of Physics and Astronomy, Ghent University, Proeftuinstraat 86, B-9000 Gent, Belgium
}
\abstract{
We present a Regge-inspired effective-Lagrangian framework for charged-kaon photoproduction from the deuteron.
Quasi-free kaon production is investigated using the Regge-plus-resonance elementary operator within the non-relativistic plane-wave impulse approximation.
The Regge-plus-resonance model was developed to describe photoinduced and electroinduced kaon production off protons
and can be extended to strangeness production off neutrons. 
The non-resonant contributions to the amplitude are modelled in terms of $K^+(494)$ and $K^{\ast+}(892)$ Regge-trajectory exchange in the $t$-channel. 
This amplitude is supplemented with a selection of $s$-channel resonance-exchange diagrams.
We investigate several sources of theoretical uncertainties on the semi-inclusive charged-kaon production cross section. 
The experimental error bars on the photocoupling helicity amplitudes turn out to put severe limits on the predictive power 
when considering quasi-free kaon production on a bound neutron.
} 
\maketitle
%
%
%

\section{Introduction}
\label{sec:PVC_introduction}

The electromagnetic~(EM) production of pseudoscalar me\-sons is the subject of intense experimental and theoretical research
for it gives access to the properties of the nucleon's excited states.
As such, these reactions provide invaluable input to our attempts to understand the strong interaction in the confinement regime. 
Associated strange\-ness production yields particularly interesting information since it involves the rearrangement of a quark-antiquark pair in the nucleon's sea.
In order to extract the masses, widths and transition form factors of nucleon and delta resonances, dynamical reaction models are required.
In spite of the extensive $p(\gamma^{(\ast)},K)Y$ data set obtained in re\-cent years, different analyses have not led to an unambiguous outcome.

Disentangling the multitude of over\-lap\-ping re\-so\-nan\-ces has proven to be challenging.
Moreover, the smooth energy dependence of the measured observables hints at a dominant role for the background, i.e.\ non-resonant, processes.
Hence, the latter are a critical ingredient of the reaction dynamics.
The Regge-plus-resonance approach to kaon production seeks to decouple the determination of the coupling constants for the background and the resonant diagrams.
Using high-energy data, we fix the non-resonant contributions to the reaction amplitude, modelled in terms of $t$-channel Regge-trajectory exchanges.
In the resonance region, this amplitude is augmented with resonance-ex\-change diagrams in the $s$-channel.
In this way, we arrive at a fair and economical description of EM kaon production data from threshold up to 
$E_{\gamma}^{\text{lab}}=16\,\text{GeV}$~\cite{RPRlambda,RPRsigma,RPRelectro,RPRneutron}.

While most research has hitherto focussed on kaon production off free protons, 
it pays to consider the same reaction on more complex targets, such as the deuteron. 
We see three major reasons to extend the Regge-plus-resonance formalism to quasi-free production on the deuteron.
First, owing to the deuteron's weak binding, it is ideally suited as an effective neutron target.
Therefore, kaon production on the deuteron gives access to the elementary $n(\gamma,K)Y$ reaction process.
Second, by comparing reactions off free and bound protons, 
our understanding of nuclear-medium effects is put to the test.
Finally, when focussing on kinematic regions where one expects important contributions from hyperons rescattering off the spectator nucleon,
allows one to gain access to the elusive hyperon-nucleon interaction.

In this contribution, we present predictions for the quasi-free production of charged kaons on the deuteron.
In the next section, we recapitulate the Regge-plus-resonance model for kaon photoproduction on both proton and neutron targets. 
Subsequently, we outline how this elementary operator can be embedded in the nuclear medium.
Sect.~\ref{sec:PVC_results} features a selection of results for the semi-inclusive quasi-free differential cross section.
Finally, we state our conclusions.

\section{Production on the free nucleon}
\label{sec:PVC_nucleon}

It is customary to study kaon photoproduction within an ef\-fec\-tive-Lagrangian model at tree-level. 
These analyses either focus exclusively on the kaon production channel or include it in a full coupled-channels approach.
In any case, the parametrisation of the non-resonant contributions to the production amplitude is a crucial and troublesome ingredient of the reaction dynamics.
The regularisation of the Born terms with ad hoc form factors and the subsequent requirement to restore gauge invariance 
implies an important model dependence for the extracted resonance information~\cite{Stijn3bckgrModels,DaveGeneticAlg}.
This prompts for an independent determination of the resonant and non-resonant contributions to the reaction amplitude.

At sufficiently high energies, hadrons are no longer the appropriate degrees of freedom and the isobar model breaks down.
Through Regge phenomenology, however, we are able to hold on to the effective-Lagrangian approach whilst ensuring the correct high-energy behaviour. 
Guidal et al.~\cite{PHDguidal,GuidalPhotoProdKandPi} demonstrated that the available high-energy kaon photoproduction data can be adequately understood 
by modelling the reaction via the exchange of linear $K^+(494)$ and $K^{\ast+}(892)$ Regge trajectories in the $t$-channel.
For each trajectory's contribution we determine the Feynman diagram for the exchange of the trajectory's first materialisation 
and subsequently replace the standard Feynman ${(t-m^2_{K^{(\ast)+}})^{-1}}$ propagator by the corresponding Regge propagator
\begin{subequations} \label{eq:PVC_reggeprop}
\begin{eqnarray}
\mathcal{P}^{K^+(494)}_{\text{Regge}} &= 
&\left(\frac{s}{s_0}\right)^{\alpha_{K^+}(t)}
\frac{
\left\lbrace \begin{array}{c}1\\e^{-i\pi\alpha_{K^+}(t)}\end{array} \right\rbrace
}{\sin\bigl(\pi\alpha_{K^+}(t)\bigr)} \; 
\frac{\pi \alpha'_{K^+}}{\Gamma\bigl(1+\alpha_{K^+}(t)\bigr)}  \,,\\
\mathcal{P}^{K^{\ast +}(892)}_{\text{Regge}} &= &
\left(\frac{s}{s_0}\right)^{\alpha_{K^{\ast+}}(t)-1} 
\frac{
\left\lbrace \begin{array}{c}1\\e^{-i\pi\alpha_{K^{\ast+}}(t)}\end{array} \right\rbrace
}{\sin\bigl(\pi\alpha_{K^{\ast+}}(t)\bigr)} \; 
\frac{\pi \alpha'_{K^{\ast+}}}{\Gamma\bigl(\alpha_{K^{\ast+}}(t)\bigr)} \,,
\end{eqnarray}
\end{subequations}
with $s_0=1\,\text{GeV}^2$, $\alpha_{K^+}(t) = 0.70 \ (t-m_{K^+}^2)$ and $\alpha_{K^{\ast+}}(t) = 1 + 0.85 \ (t-m_{K^{\ast+}}^2)$, 
when $t$ and $m_{K^{(\ast)+}}^2$ are expressed in units of $\text{GeV}^2$. 
As can be appreciated from Eq.~(\ref{eq:PVC_reggeprop}), a phase option, either constant ($1$) or rotating ($e^{-i\pi\alpha(t)}$),
remains for the Regge propagators. 
A rotating phase for both trajectories is best suited to describe $\Lambda$ production~\cite{RPRlambda},
while the $K\Sigma$ channel prefers a rotating (constant) phase for the $K^+(494)$ ($K^{\ast+}(892)$) trajectory~\cite{RPRsigma}.

A crucial constraint for the kaon-production amplitude is gauge invariance. 
It is well-known that the $t$-channel Born diagram by itself does not conserve electric charge. 
In Ref.~\cite{GuidalPhotoProdKandPi}, an elegant recipe to correct for this was outlined. 
Adding the electric part of a Reggeized $s$-channel Born diagram ensures that the amplitude is gauge invariant.

The Regge model's amplitude can be interpreted as the asymptotic form of the full amplitude for large $s$ and small $|t|$. 
Owing to the $t$-channel dominance and the absence of a prevailing resonance, 
the Reggeized background can account for the gross features of the kaon-production data within the resonance region~\cite{RPRsigma,GuidalUpdate}. 
Near threshold, the energy dependence of the measured differential cross sections exhibits structure which hints at the presence of resonances. 
These are incorporated by supplementing the background with a number of resonant $s$-channel diagrams using standard tree-level Feynman diagrams. 
This approach was coined Regge-plus-resonance~(RPR) and results in the following structure for the elementary production operator:
\begin{eqnarray}\label{eq:PVC_RPRop}
&&\hat{J}_{\text{elem}}\left(\gamma p\rightarrow K^+Y\right) = \nonumber\\
 & & \hat{J}_{\text{Regge}}^{K^{+}(494)}
   +\hat{J}_{\text{Regge}}^{K^{\ast+}(892)} 
 +\hat{J}_{\text{Feyn}}^{\text{Born-s,elec}}
     \times \mathcal{P}_{\text{Regge}}^{K^{+}(494)}
     \times \left(t-m_K^2\right)\nonumber\\
 && + \sum_{N^{\ast}} \hat{J}^{N^{\ast}}_{\text{Feyn}}
    + \sum_{\Delta^{\ast}} \hat{J}^{\Delta^{\ast}}_{\text{Feyn}}\,.    
\end{eqnarray}

In the $K^+\Lambda$ and $K^+\Sigma^0$ production channels, sufficient data is available to constrain the free parameters of the production operator.
The third possible final state containing a charged kaon ($K^+\Sigma^-$), on the other hand, suffers from a lack of data 
and extrapolations are unavoidable.
Within the RPR model the $n(\gamma,K^+)\Sigma^-$ and $p(\gamma,K^+)\Sigma^0$ reactions can be related to each other.
As was outlined in Ref.~\cite{RPRneutron}, it suffices to convert the coupling constants which feature in the interaction Lagrangians.
In the strong interaction vertex, we assume SU(2) isospin symmetry to be exact. 
Taking the hadronic couplings proportional to the Clebsch-Gordan coefficients, we obtain the following relations
\begin{subequations}\label{eq:PVC_strongIsospin}
 \begin{eqnarray}\textstyle
 g_{K^{(\ast)+}\Sigma^-n} &= &\sqrt{2}\,g_{K^{(\ast)+}\Sigma^0p}\,,\\
 g_{K^{(\ast)+}\Sigma^-N^{\ast0}} &= &\sqrt{2}\,g_{K^{(\ast)+}\Sigma^0N^{\ast+}}\, .
 \end{eqnarray}
\end{subequations}

\begin{table}
\caption{%
The ratio of the EM coupling constants to proton and neutron for selected nucleon resonances 
obtained with Eq.~(\ref{eq:PVC_EMratio12}) or Eq.~(\ref{eq:PVC_EMratio32}). 
The listed values are obtained using photocoupling helicity amplitudes from SAID analysis SM95~\cite{SAID96}.
No experimental information exists for the $N(1900)P_{13}$,
therefore we consider a broad range.%
}
\label{tab:PVC_helamp}
\begin{tabular}{lr@{$\,\pm\,$}rr@{$\,\pm\,$}rr@{$\,\pm\,$}r}
\hline\noalign{\smallskip}
Resonance  & \multicolumn{2}{c}{$\frac{\kappa_{ {N^*}n}}{ \kappa_{ { N^*}p}}$} 
           & \multicolumn{2}{c}{$\frac{\kappa_{ {N^*}n}^{\left( 1 \right)}}{ \kappa_{ { N^*}p}^{\left( 1 \right)}}$}
           & \multicolumn{2}{c}{$\frac{\kappa_{ { N^*}n}^{\left( 2 \right)}}{ \kappa_{ {N^*}p}^{\left( 2 \right)}}$} \\ 
\noalign{\smallskip}\hline\noalign{\smallskip}
$N(1650)S_{11}$ &$-0.22$ &$0.07$ &\multicolumn{2}{c}{$-$} &\multicolumn{2}{c}{$-$}  \\ 
$N(1710)P_{11}$ &$-0.29$ &$2.23$ &\multicolumn{2}{c}{$-$} &\multicolumn{2}{c}{$-$} \\ 
$N(1720)P_{13}$ &\multicolumn{2}{c}{$-$} &$-0.38$ &$2.00$ &$-0.50$ &$1.08$ \\ 
$N(1900)P_{13}$ &\multicolumn{2}{c}{$-$} &$ 0.00$ &$2.00$ &$ 0.00$ &$2.00$ \\ 
\noalign{\smallskip}\hline
\end{tabular}
\end{table}

In the Regge model, the coupling constants at the EM vertex do not change. 
The procedure to restore gauge invariance, however, is slightly altered. 
This time we need to include the electric part of a Reggeized $u$-channel Born diagram.
Converting the EM coupling constant of the resonant contributions requires knowledge of the photocoupling helicity amplitudes $\mathcal{A}^N_{J}$,
which are extracted from pion production reactions, amongst others.
It can be shown that the ratio of a resonance's EM couplings constants to neutron and proton is given by
\begin{equation}\label{eq:PVC_EMratio12}
 \frac{\kappa_{ {
N^*}n}}{ \kappa_{ { N^*}p}} = \frac{\mathcal{A}^n_{1/2}}{\mathcal{A}^p_{1/2}}\,,
\end{equation}
or
\begin{subequations}\label{eq:PVC_EMratio32}
\begin{eqnarray}
 \frac{\kappa_{ {N^*}n}^{\left( 1 \right)}}{ \kappa_{ { N^*}p}^{\left( 1 \right)}} 
 &=
 &\frac{ \sqrt{3} \mathcal{A}^n_{1/2} \pm \mathcal{A}^n_{3/2}}
      {\sqrt{3} \mathcal{A}^p_{1/2} \pm \mathcal{A}^p_{3/2}}\\
 \frac{\kappa_{ { N^*}n}^{\left( 2 \right)}}{ \kappa_{ {N^*}p}^{\left( 2 \right)}} 
 &= 
 &\frac{ \sqrt{3} \mathcal{A}^n_{1/2} - \frac{m_p}{m_{ N^*}} \mathcal{A}^n_{3/2}}
      {\sqrt{3} \mathcal{A}^p_{1/2} - \frac{m_p}{m_{ N^*}} \mathcal{A}^p_{3/2}}\,.
\end{eqnarray}
\end{subequations}
for spin-$1/2$ or spin-$3/2$ resonances respectively~\cite{RPRneutron}.
In Table~\ref{tab:PVC_helamp} we list the conversion factors for the resonances relevant to our model
employing the helicity amplitudes extracted in the SAID analysis SM95~\cite{SAID96}.
It is immediately clear that the ratios have considerable error bars.
Moreover, no information is available for the $N(1900)P_{13}$ resonance.
Therefore, we allow the ratios of its magnetic transition moments, $\kappa_{ { N^*}n}^{\left( 1,2 \right)}/ \kappa_{ {N^*}p}^{\left( 1,2 \right)}$, 
to vary between $-2$ and $+2$.

In Refs.~\cite{RPRlambda,RPRsigma}, we presented results for the $p(\gamma,K^+)\Lambda$ and $p(\gamma,K^+)\Sigma^0$ reactions. 
The free parameters of the Regge model were fitted to the available high-energy data. 
Next, the coupling constants of the resonance-exchange diagrams were determined by 
optimizing the model against photoproduction data in the resonance region ($E_{\gamma}^{\text{lab}}\leq2.5\,\text{GeV}$),
while the parameters of the background contributions are frozen.
The resulting models allow for a good description of the available data
and exhibit robustness when compared to newly measured polarisation observables~\cite{CLASbeamrec,LleresBeamRec}.
In addition, the RPR model has demonstrated its high predictive power through its nice agreement with electroproduction data~\cite{RPRelectro}.
When the RPR model was compared to the limited $n(\gamma,K^+)\Sigma^-$ data set, it was observed that the Regge model produces satisfactory predictions. 
The predictive power of the full RPR model, on the other hand, is limited by the experimental error bars of the helicity amplitudes~\cite{RPRneutron}.

\section{Quasi-free production on the deuteron}
\label{sec:PVC_deuteron}

In the previous section, we have presented the elementary charged-kaon production operator in the Regge-plus-resonance approach.
Now, we will sketch the formalism to embed this operator in the deuteron.
If one assumes the conventions of Peskin and Schroeder~\cite{PeskinAndSchroeder}, the expression for the $^2H(\gamma,KY)N$ cross section is given by
\begin{eqnarray}\label{eq:PVC_diffXsection}
 d\sigma &= &\frac{1}{|\vec{v}_{\gamma}-\vec{v}_D|}\frac{1}{2E_{\gamma}2E_D} 
	   \frac{d^3\vec{p}_K}{2E_K(2\pi)^3}
           \frac{d^3\vec{p}_Y}{2E_Y(2\pi)^3}
           \frac{d^3\vec{p}_N}{2E_N(2\pi)^3} \nonumber\\
	   &&\times(2\pi)^4\delta^{(4)}(p_{\gamma}+p_D-p_K-p_Y-p_N) \nonumber\\
           &&\times \frac{1}{6} \overline{\sum} 2
           \left|\left\langle \vec{p}_{K} \vec{p}_Y \vec{p}_{N} \lambda_K \lambda_Y \lambda_N \right| 
                  \epsilon_{\lambda_{\gamma}}\cdot \hat{J}_{\text{elem}}
                 \left| \vec{p}_D \lambda_D  \right\rangle \right|^2\,.
\end{eqnarray}
Here $\epsilon_{\lambda_{\gamma}}$ is the photon's polarisation vector and $p_{\gamma}$, $p_D$, $p_K$, $p_Y$ and $p_N$ are the momenta of the photon, deuteron, kaon, hyperon and nucleon respectively.
Their spin projections are denoted by $\lambda_{\gamma}$, $\lambda_D$, $\lambda_Y$ and $\lambda_N$, 
and $1/6\overline{\sum}$ implies a summation (averaging) over the spins of the initial (final) state particles.
To obtain the semi-inclusive kaon-production cross section, we integrate over the four-momenta of the hyperon and nucleon in their centre-of-mass frame.
All other quantities are evaluated in the laboratory frame. We have
\begin{eqnarray}\label{eq:PVC_incXsection}
 \frac{d^3\sigma}{dp_Kd\Omega_K} &= 
 &\int d\Omega_Y^{\ast}
  \frac{1}{32(2\pi)^5}\frac{|\vec{p}_Y^*||\vec{p}_K|^2}{M_D E_{\gamma}^{\text{lab}} E_K W_{YN}} \\
 &\times &\frac{1}{6} \overline{\sum}
           2\left|\left\langle \vec{p}_{K} \vec{p}_Y \vec{p}_{N} \lambda_K \lambda_Y \lambda_N \right| 
                  \epsilon_{\lambda_{\gamma}}\cdot \hat{J}_{\text{elem}}
           \left| \vec{p}_D \lambda_D  \right\rangle \right|^2\, ,\nonumber
\end{eqnarray}
with $W_{YN}$ the invariant mass of the hyperon-nucleon system.

When evaluating the matrix element in Eqs.~(\ref{eq:PVC_diffXsection}) and (\ref{eq:PVC_incXsection}), we invoke the impulse approximation~(IA),
which allows to write the interaction Hamiltonian as an incoherent sum of one-body current operators. 
Our notation tacitly implies that the elementary operator (see Eq.~(\ref{eq:PVC_RPRop})) only acts on particle `1' inside the deuteron.
Were $\hat{J}_{\text{elem}}$ to act on particle `2', the amplitude would be the same. Therefore, we multiply the matrix element by two.

The kinematics and the elementary operator are treated in a fully relativistic manner.
It has been shown that the use of a relativistic wave function has negligible effects on the quasi-free cross section
as long as the momentum of the final-state nucleon is small compared to its mass ~\cite{AdelseckWright}.
Therefore, we have opted for a non-relativistic deuteron wave function.
The deuteron state can be decomposed as
\begin{eqnarray}\label{eq:PVC_Dstate}
\left|\vec{p}_D\lambda_D\right\rangle &= 
 &\sqrt{2E_D \left(2\pi \right)^3}
  \int d^3\vec{p}'_N \sqrt{\frac{E_N}{E'_N}}
  \delta^{(3)}\left(\vec{p}_D-\vec{p}_N-\vec{p}'_N\right) \nonumber\\
 &\times &\frac{1}{\sqrt{2}}\left(
                                  \left| {\textstyle m^I_1=\frac{1}{2},m^I_2=-\frac{1}{2}}\right\rangle 
                                 -\left| {\textstyle m^I_1=-\frac{1}{2},m^I_2=\frac{1}{2}}\right\rangle 
    \right)\nonumber\\
 &\times &\sum_{\lambda'_N}\sum_{m_S}\sum_{L=0,2}\sum_{m_L} 
          \left| \vec{p}'_N \lambda'_N \right\rangle
          i^L
          u_L(|\vec{p}'_N|)
          Y_{Lm_L}\left(\hat{\vec{p}}'_N\right)  \nonumber\\  
 &\times & \left\langle {\textstyle \frac{1}{2} \lambda_N ; \frac{1}{2} \lambda'_N} \right|
           \left. {\textstyle \frac{1}{2}\frac{1}{2},1 m_S} \right\rangle
           \left\langle {\textstyle Lm_L ; 1 m_S} \right|
           \left. {\textstyle L1,1 \lambda_D} \right\rangle\,,
\end{eqnarray}
with $p'_N$ ($\lambda'_N$) the four-momentum (spin projection) of the struck nucleon inside the deuteron.
The $Y_{Lm_L}\left(\theta,\phi\right)$ are spherical harmonics 
and $u_L(p)$ are the $s$-wave and $d$-wave components of the non-relativistic deuteron wave function.
For completeness, we have also written down the isospin component of the deuteron state.

Because of the Dirac delta function in Eq.~(\ref{eq:PVC_Dstate}), the struck nucleon is never on its mass shell, i.e.\ $p^{'2}_N<m^2$.
One can deal with this by either using $p^{'2}_N$ as an effective mass in the elementary operator
or by forcing the struck nucleon on its mass shell and thus violating conservation of energy in the $Dnp$-vertex.
In the forthcoming section, the influence of these different prescriptions will be investigated.

\section{Results}
\label{sec:PVC_results}	

Within the plane-wave impulse approximation, the photoproduction of a charged kaon from a deuteron target can proceed in three different elementary ways.
For two of these, $^2H(\gamma,K^+)\Lambda n$ and $^2H(\gamma,K^+)\Sigma^0n$, the incoming photon interacts with a bound proton inside the nucleus.
The elementary operators for these reactions are well constrained by the existing data.
This is in sharp contrast with the case of neutral-kaon production, where no experimental information to fix the RPR model is available.
For this reason, we will only present results for the charged-kaon production cross section.

\begin{figure*}[!htb]
\centering
\includegraphics[width=\columnwidth]{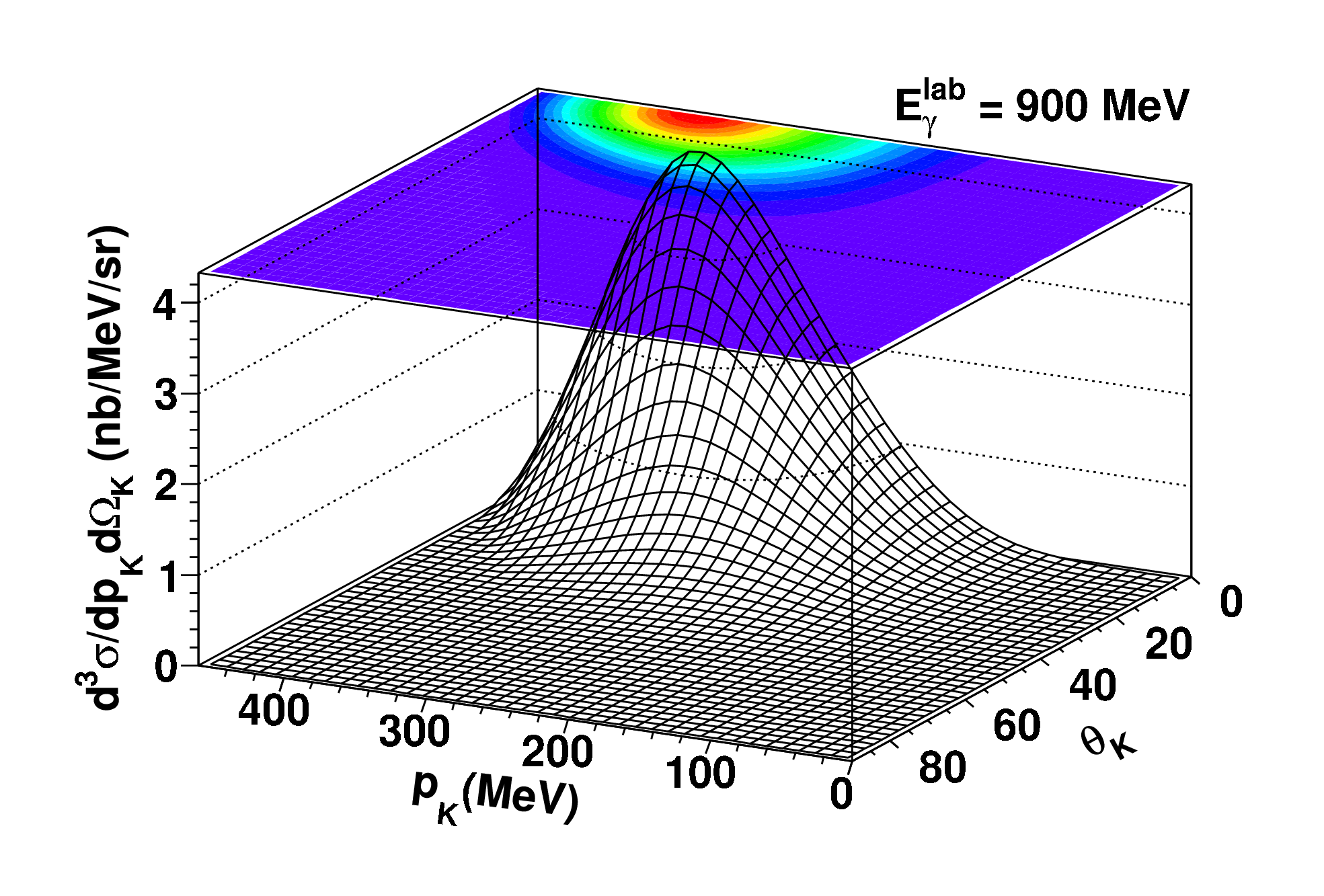}
\includegraphics[width=\columnwidth]{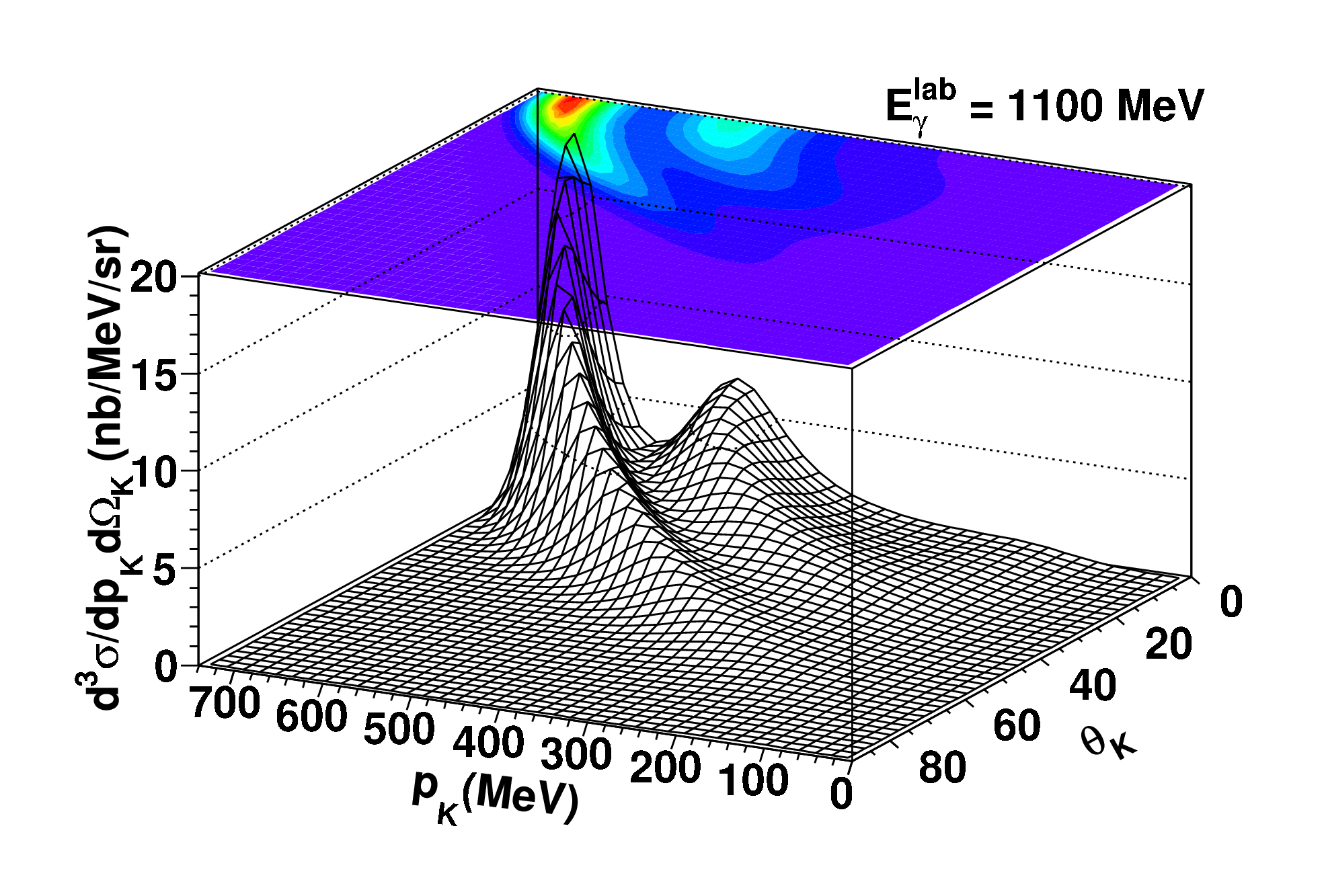}\\
\includegraphics[width=\columnwidth]{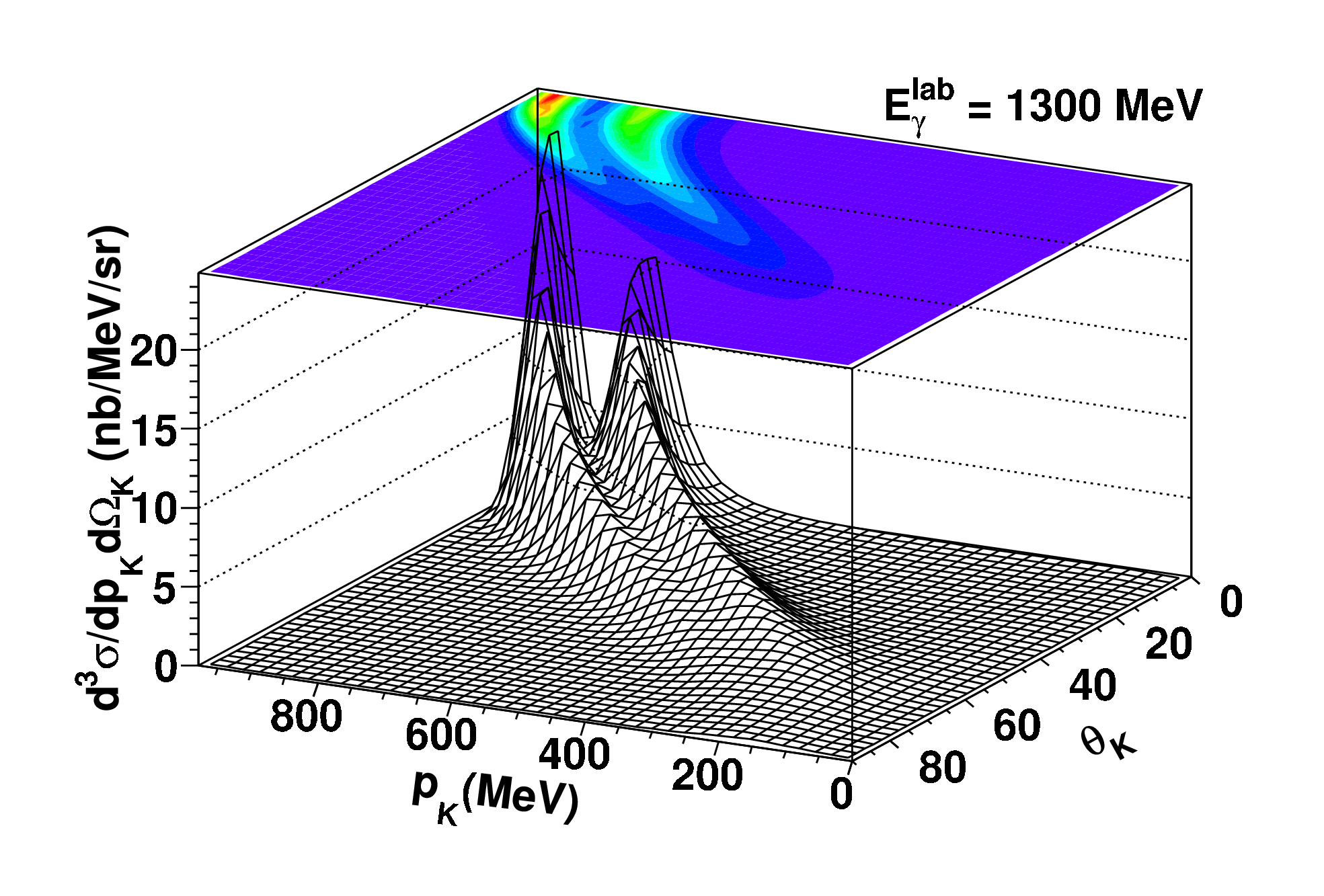}
\includegraphics[width=\columnwidth]{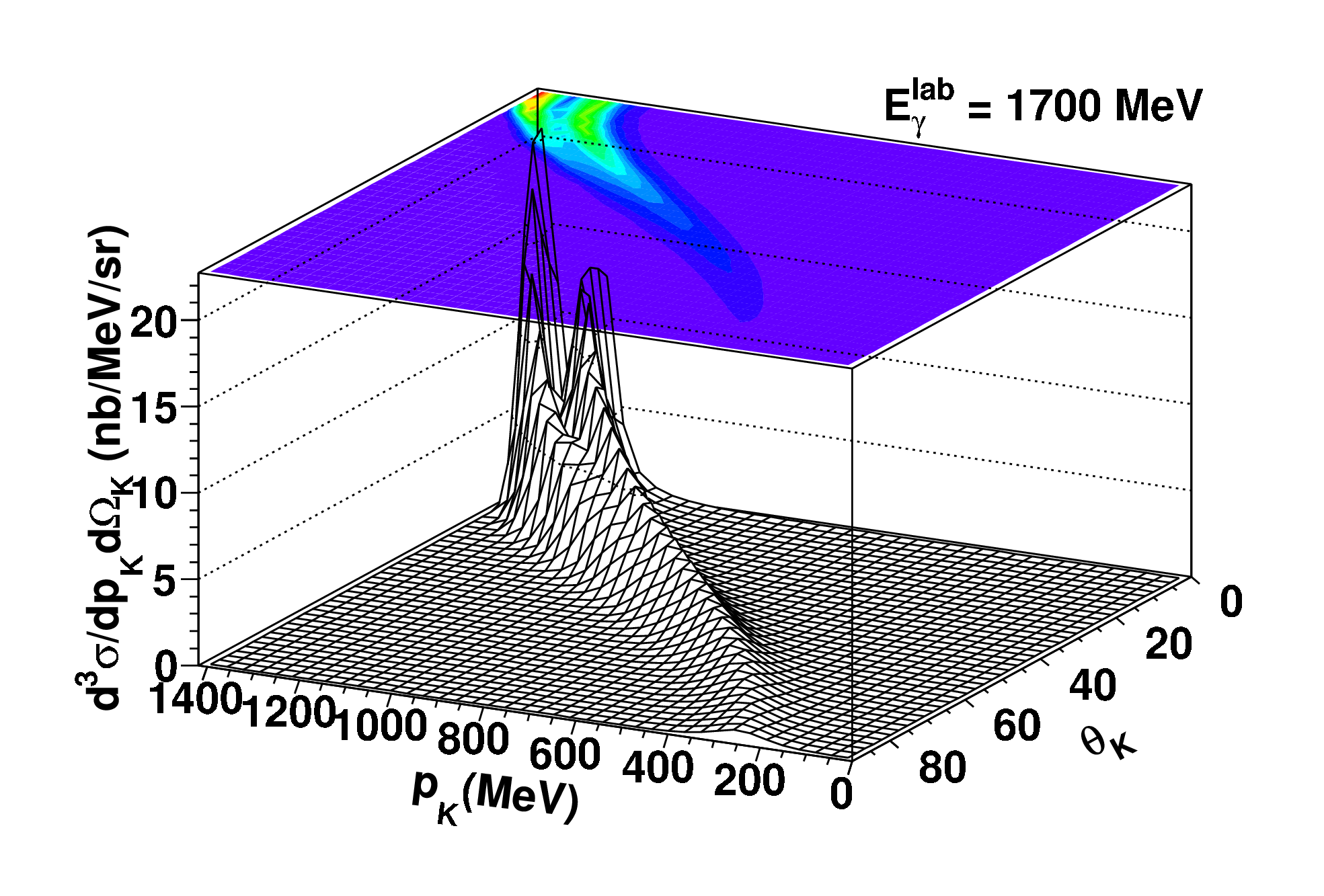}\\
\includegraphics[width=\columnwidth]{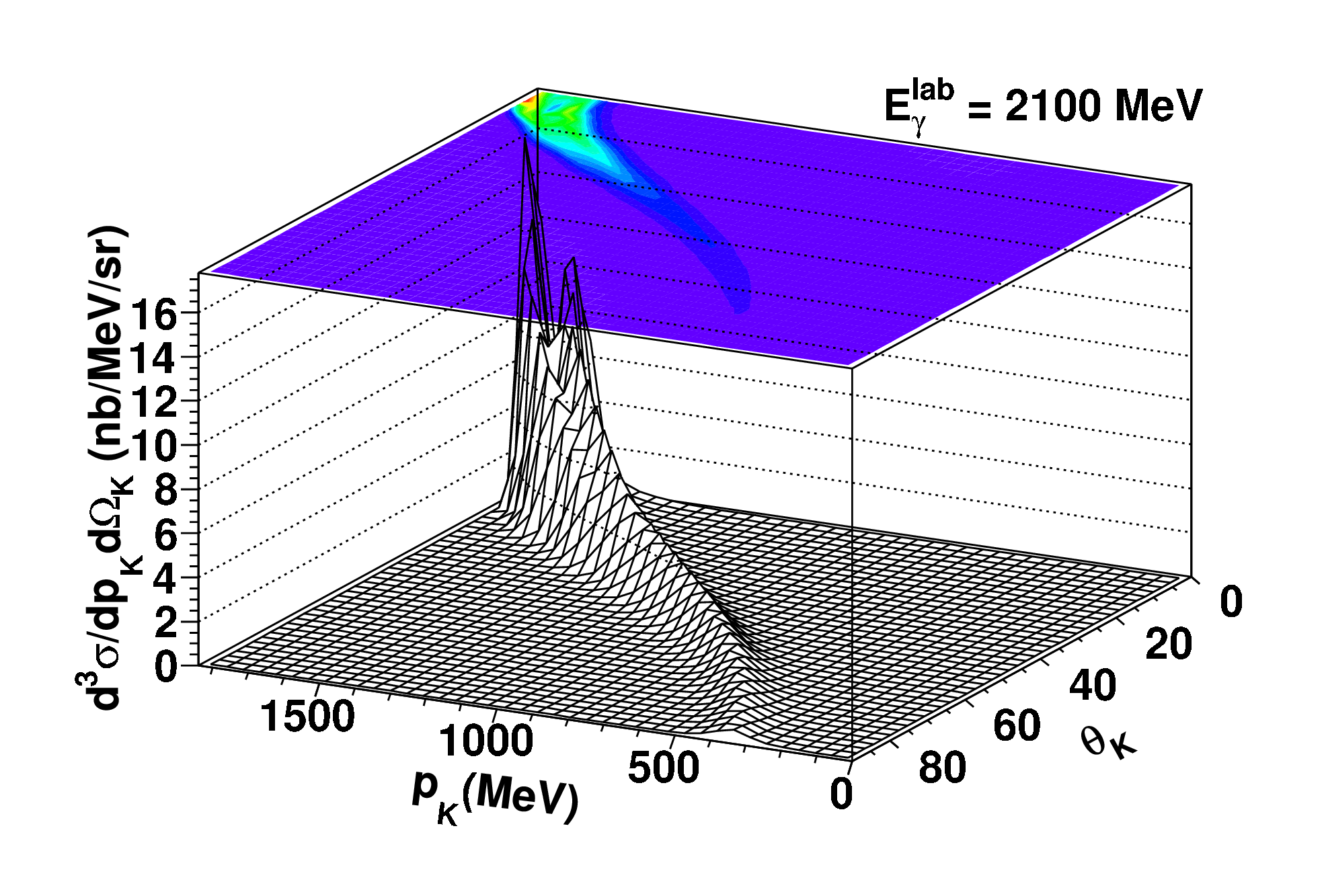}
\includegraphics[width=\columnwidth]{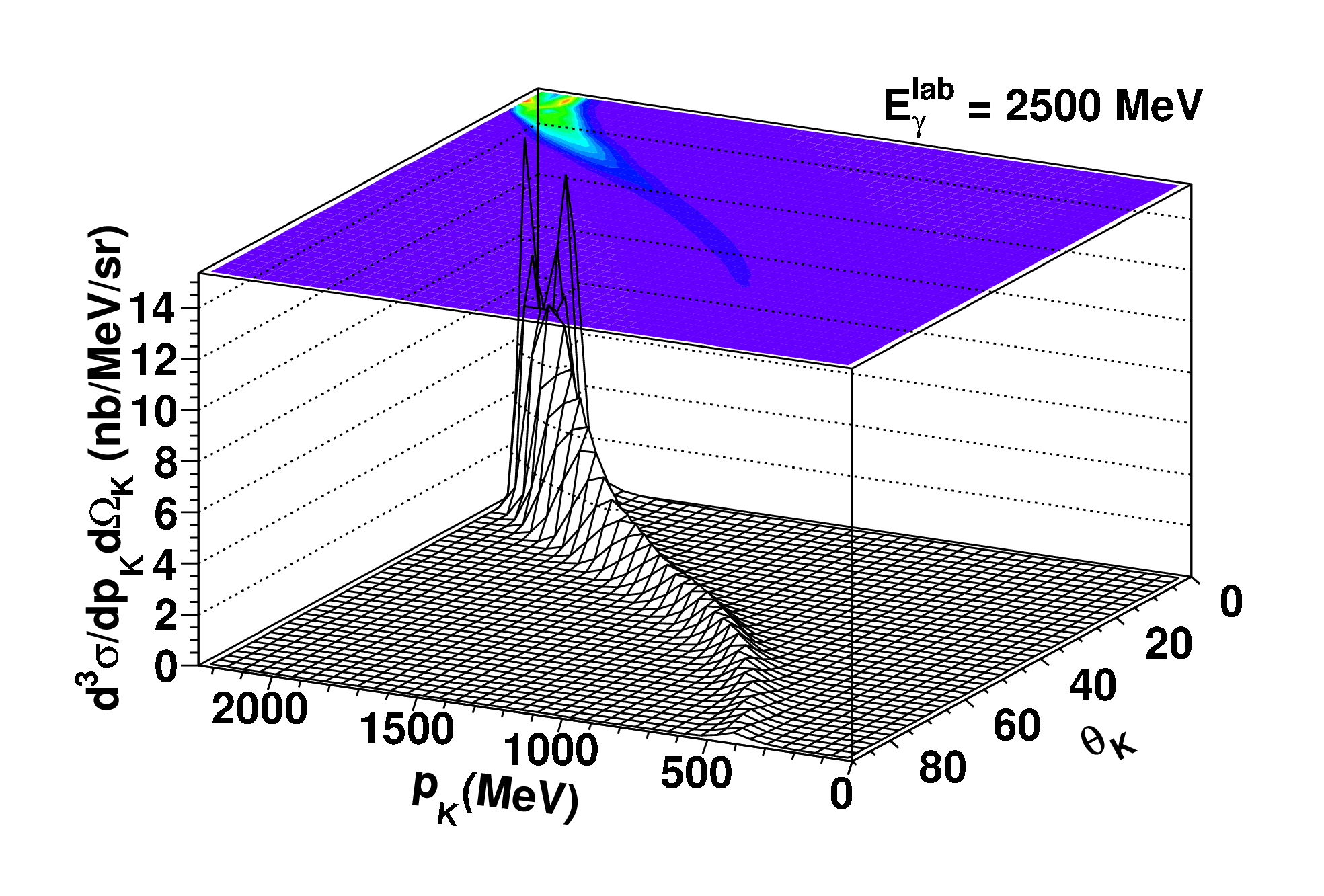}
\caption{The semi-inclusive ${^2H}(\gamma,K^+)YN$ differential cross section 
using the full RPR operator
as a function of the kaon momentum $p_K$ and scattering angle $\theta_K$ in the laboratory system 
at six different photon energies (from top-left to bottom-right $E_{\gamma}^{\text{lab}}=900,1100,1300,1700,2100\,\text{and}\,2500\,\text{MeV}$).}
\label{fig:PVC_semi-inclusive3D}
\end{figure*}

In Fig.~\ref{fig:PVC_semi-inclusive3D}, the threefold differential cross section is shown as a function of the kaon's momentum and scattering angle in the laboratory frame.
These results are obtained with the RPR current operator and use the struck nucleon's four-mo\-men\-tum squared $p^{'2}_N$ as an effective mass.
The employed deu\-ter\-on wave function is generated by the charge-dependent Bonn potential (CD-Bonn)~\cite{CDBonnWF}.
The shape of the semi-inclusive cross section is mainly determined by the deuteron's momentum distribution, which peaks at vanishing relative momentum.
The cross section is largest when kaons are created along the direction of the incoming photon.
At all but the lowest photon energy, the cross section consists of two modi.
The one at the highest kaon momenta corresponds to $\Lambda$ production, the second to $\Sigma^0$/$\Sigma^-$ production.
As the photon energy increases, the gap between both ridges becomes smaller and disappears at $E_{\gamma}^{\text{lab}}\approx2\,\text{GeV}$.

\begin{figure}[!ht]
\centering
\includegraphics[width=\columnwidth]{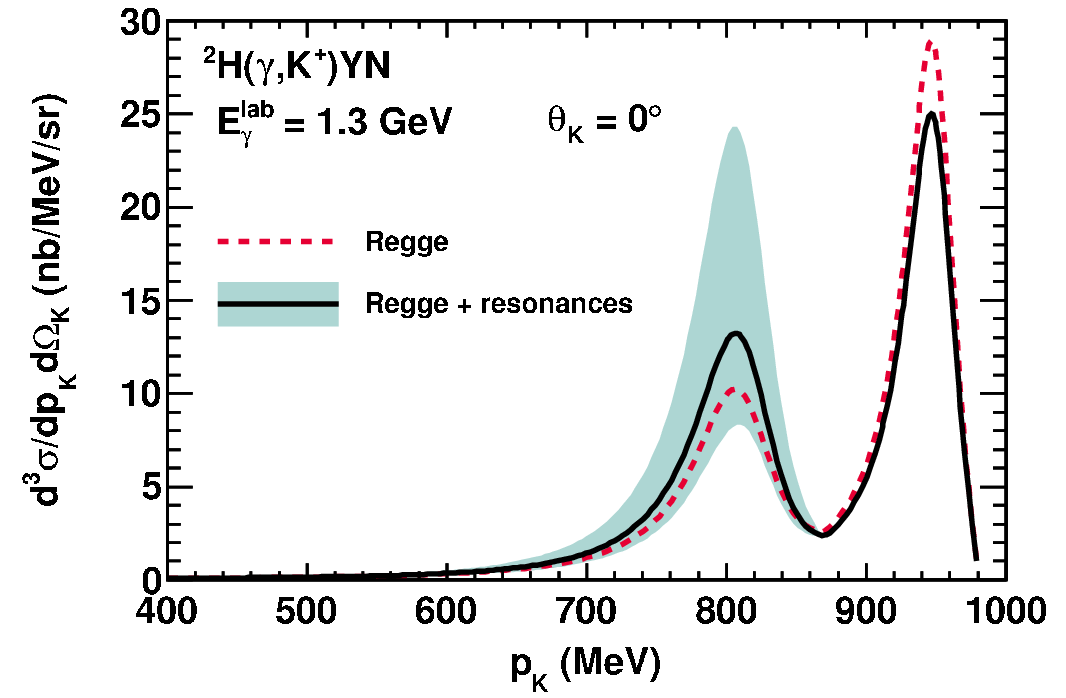}
\caption{The threefold differential ${^2H}(\gamma,K^+)YN$ cross section 
as a function of the kaon momentum $p_K$ at forward-scattering angle $\theta_K$
and photon energy $E_{\gamma}^{\text{lab}}=1300\,\text{MeV}$.
The dashed curve shows the contribution of the Reggeized background,
whereas the solid curve also includes the $s$-channel resonant contributions of the full RPR amplitude.
The shaded area takes the uncertainties of the adopted helicity amplitudes into account. 
These uncertainties are listed in Table~\ref{tab:PVC_helamp}.
}
\label{fig:PVC_helampError}
\end{figure}

The semi-inclusive cross section at forward kaon scattering angle and at $E_{\gamma}^{\text{lab}}=1300\,\text{MeV}$ is shown in Fig.~\ref{fig:PVC_helampError},
using the current operator of both the Regge as well as the more complete RPR model. 
The shape and size of the semi-inclusive cross section are compatible with the results presented in Fig.~6 of Ref.~\cite{Yamamura},
even though a different wave function and elementary operator were used.
Interestingly, the resonant contributions to the production operator add strength to the $\Sigma$ peak,
whereas they produce destructive interference with the non-resonant diagrams in the $\Lambda$-production peak.

Several sources of uncertainties may affect the calculated cross sections. 
Besides the CD-Bonn wave functions, we have also applied wave functions obtained with the Paris~\cite{ParisPot} and Nijmegen-III~\cite{NijmegenPotential} potentials.
This does not have any impact on the results.
Likewise, the two above-mentioned prescriptions to deal with the off-shell character of the nucleons provide similar results.
This can be understood by realising that the major fraction of the semi-inclusive cross section's strength stems from those regions in phase space
where the struck nucleon is marginally off its mass shell.

In Ref.~\cite{RPRneutron} it became apparent to what extent the experimental errors on the photocoupling helicity amplitudes of established resonances
affect the predictability of the $n(\gamma,K^+)\Sigma^-$ cross section. 
In Fig.~\ref{fig:PVC_helampError}, the effect of these uncertainties is indicated by the shaded area.
The induced deviations are important.
In the $\Sigma$-production peak, variations of the order of two are possible in the predicted cross sections.
Measuring the threefold differential cross section, even with limited accuracy, 
would therefore allow to put stringent constraints on the resonances' helicity amplitudes.

\section{Conclusions}
\label{sec:PVC_conclusions}

We have presented results for quasi-free charged-kaon photoproduction from the deuteron within a non-relativistic plane-wave impulse approximation.
For the elementary process, we employ the Regge-plus-resonance model, which has been optimised against recent photoproduction data.
We have shown that the adopted deuteron wave function and the prescription to treat the off-shell nature of the struck nucleon
have an insignificant effect on the calculated threefold differential cross sections.
As in the elementary $n(\gamma,K^+)\Sigma^-$ channel, the experimental errors on the resonances' helicity amplitudes are the most important source of uncertainties.

\begin{acknowledgement}
This work was supported by the Research Foundation - Flanders (FWO) and the research council of Ghent University.
\end{acknowledgement}

\bibliographystyle{epj}
\bibliography{bibliography}

\end{document}